\documentclass[onecolumn,unpublished]{quantumarticle}
\usepackage{amsmath}
\usepackage{amssymb}
\usepackage{amsthm}
\usepackage{amsfonts}
\usepackage[caption=false]{subfig}
\usepackage[colorlinks]{hyperref}
\usepackage[all]{hypcap}
\usepackage{tikz}
\usepackage{float}
\usepackage{listings}
\usepackage{relsize}
\usepackage{color,soul}
\usepackage[utf8]{inputenc}
\usepackage{capt-of}
\usepackage[numbers]{natbib}
\usepackage[T1]{fontenc}  

\usetikzlibrary{decorations.pathreplacing}

\renewcommand\thesection{\arabic{section}}

\theoremstyle{definition}

\theoremstyle{definition}

\theoremstyle{definition}

\newcommand{\eq}[1]{\hyperref[eq:#1]{Equation~\ref*{eq:#1}}}
\renewcommand{\sec}[1]{\hyperref[sec:#1]{Section~\ref*{sec:#1}}}
\DeclareRobustCommand{\app}[1]{\hyperref[app:#1]{Appendix~\ref*{app:#1}}}
\newcommand{\fig}[1]{\hyperref[fig:#1]{Figure~\ref*{fig:#1}}}
\newcommand{\tbl}[1]{\hyperref[tbl:#1]{Table~\ref*{tbl:#1}}}
\newcommand{\theoremref}[1]{\hyperref[theorem:#1]{Theorem~\ref*{theorem:#1}}}
\newcommand{\definitionref}[1]{\hyperref[definition:#1]{Definition~\ref*{definition:#1}}}

\DeclareFixedFont{\ttb}{T1}{txtt}{bx}{n}{12}
\DeclareFixedFont{\ttm}{T1}{txtt}{m}{n}{12}
\definecolor{deepblue}{rgb}{0,0,0.5}
\definecolor{deepred}{rgb}{0.6,0,0}
\definecolor{deepgreen}{rgb}{0,0.5,0}
\newcommand\pythonstyle{\lstset{
language=Python,
basicstyle=\ttm,
otherkeywords={self,controlledby,with,quint,let,carryinto,store},
keywordstyle=\ttb\color{deepblue},
emph={measure,__init__},
emphstyle=\ttb\color{deepblue},
stringstyle=\color{deepgreen},
showstringspaces=false
}}
\lstnewenvironment{python}[1][]
{\pythonstyle\lstset{#1}}{}

\DeclareFixedFont{\ttbq}{T1}{txtt}{bx}{n}{8}
\DeclareFixedFont{\ttmq}{T1}{txtt}{m}{n}{8}
\definecolor{deepblue}{rgb}{0,0,0.5}
\definecolor{deepred}{rgb}{0.6,0,0}
\definecolor{deepgreen}{rgb}{0,0.5,0}
\newcommand\qsharpstyle{\lstset{
basicstyle=\ttmq,
morekeywords={operation,LittleEndian,let,within,Unit,is,Adj,CNOT,using,Qubit,open,namespace,apply,for,in,function,Int,while,set,return,mutable,H,M,CZ,One,Zero,adjoint,body,CCNOT,Length,Chunks,new,X,Adjoint,fail,if,and,or},
keywordstyle=\ttbq\color{deepblue},
emph={measure,__init__},
emphstyle=\ttbq\color{deepblue},
stringstyle=\color{deepgreen},
morecomment=[l]{//},
commentstyle=\color{deepgreen},
showstringspaces=false,
literate={*}{{\char42}}1
         {-}{{\char45}}1
}}
\lstnewenvironment{qsharp}[1][]
{\qsharpstyle\lstset{#1}}{}

\input{qcircuit}

\title{Quantum block lookahead adders and the wait for magic states}
\date{\today}
\author{Craig Gidney}
\email{craiggidney@google.com}
\affiliation{Google Inc., Santa Barbara, California 93117, USA}

\begin{document}
\maketitle

\begin{abstract}
We improve the Toffoli count of low depth quantum adders, and analyze how their spacetime cost reacts to having a limited number of magic state factories.
We present a block lookahead adder that parallelizes across blocks of bits of size $b$, instead of over all bits.
The block lookahead adder achieves a Toffoli count of $3n + 5n/b$ for out of place addition (vs $4n$ in previous work by Thapliyal et al) and $5n + 8n/b$ for in place addition (vs $7n$ in previous work by Thapliyal et al).
The tradeoff is that the reaction depth of these circuits depends linearly on $b$, and they use additional workspace.
We estimate the spacetime volume of these adders, and adders from previous work, for various register sizes and factory counts under plausible assumptions for a large scale quantum computer based on the surface code and superconducting qubits.
\end{abstract}

\section{Introduction}

In the classical computing world, low depth adders are everywhere.
Even the original 8-bit Intel 8008 chip had a carry lookahead adder \cite{shirriff2020reverseengineer8008}.
But it's not clear if low depth adders will enjoy the same popularity in the world of fault tolerant quantum computation, because of huge implied space overheads from magic state distillation.

The surface code \cite{fowler2012surfacereview} is a leading contender for the error correcting code used by future large scale quantum computers, because it has a high threshold and low connectivity requirements.
In the surface code, non-Clifford operations like the T gate and the Toffoli gate aren't native operations.
They have to be emulated using roundabout techniques like magic state distillation \cite{bravyi2005magicstate}.
Magic state factories are expected to be expensive.
For example in \cite{gidney2019catalyzed} a factory covers hundreds of thousands of physical qubits (although new techniques are reducing the cost \cite{litinski2019magicnotcostly}).

If a quantum computer can't host enough magic state factories, a low depth adder will bottleneck waiting for magic states.
Suppose we have a quantum computer with a control system reaction time of 10 microseconds and a factory design that produces a Toffoli state roughly every 165 microseconds as in \cite{gidney2019autoccz}.
Using a ripple-carry adder and reaction limited quantum computation \cite{fowler2012timeoptimal,gidney2019autoccz}, we would need 17 magic state factories before the ripple carry process was running at top speed.
If a low depth adder takes five times more Toffolis, we need five times more factories ($17 \cdot 5 = 85$) to compensate.
Ultimately this suggests that, if we can't dedicate tens of millions of physical qubits to magic state distillation, the ripple carry adder will be faster by default.

\begin{table}
\centering
\resizebox{\linewidth}{!}{
\begin{tabular}{r|c|c|l|l|l|l|c|c|c}
Paper                                                &Place &Type                    &Toffolis                        &Reaction Depth                 &Workspace                  &Log Toffoli / Time (n=128,b=10)                                                                                                                                                                                                                                                                                                                                                                                                                                                                                                                                                                                                                                                                                                                                                                                                                                                                                                                                                                                                                                                                                                                                                                                                                                                                                                                          &V(n=100,f=10) &V(n=1000,f=100) &V(n=10000,f=1000) \\
\hline
Cuccaro (2004) \cite{cuccaro2004adder}               &in    &Ripple Carry            &$2n - 1$                        &$2n - 1$                       &$1$                        &\begin{tikzpicture}\fill[red] (0.0,-6.25e-07) -- (0.0,0.0) -- (0.0,0.0625) -- (4.98046875,0.0625) -- (4.98046875,-6.25e-07) -- cycle;\draw (0,0.5) -- (0,0) -- (4.98046875,0) -- (4.98046875,0.5); \end{tikzpicture}                                                                                                                                                                                                                                                                                                                                                                                                                                                                                                                                                                                                                                                                                                                                                                                                                                                                                                                                                                                                                                                                                                                                     &3             &63              &4200              \\
Draper et al. (2004) \cite{draper2004lookaheadadder} &in    &Carry Lookahead         &$10n - 6\lg n - 13$             &$4\lg n + 7$                   &$2n - \lg n - 1$           &\begin{tikzpicture}\fill[red] (0.0,-6.25e-07) -- (0.0,0.0) -- (0.0,0.5) -- (0.01953125,0.5) -- (0.01953125,0.4375) -- (0.0390625,0.4375) -- (0.0390625,0.5) -- (0.05859375,0.5) -- (0.05859375,0.4375) -- (0.078125,0.4375) -- (0.078125,0.375) -- (0.09765625,0.375) -- (0.09765625,0.3125) -- (0.1171875,0.3125) -- (0.1171875,0.25) -- (0.13671875,0.25) -- (0.13671875,0.125) -- (0.17578125,0.125) -- (0.17578125,0.25) -- (0.1953125,0.25) -- (0.1953125,0.3125) -- (0.21484375,0.3125) -- (0.21484375,0.375) -- (0.234375,0.375) -- (0.234375,0.4375) -- (0.25390625,0.4375) -- (0.25390625,0.5) -- (0.2734375,0.5) -- (0.2734375,0.4375) -- (0.3125,0.4375) -- (0.3125,0.5) -- (0.33203125,0.5) -- (0.33203125,0.4375) -- (0.3515625,0.4375) -- (0.3515625,0.375) -- (0.37109375,0.375) -- (0.37109375,0.3125) -- (0.390625,0.3125) -- (0.390625,0.25) -- (0.41015625,0.25) -- (0.41015625,0.125) -- (0.44921875,0.125) -- (0.44921875,0.25) -- (0.46875,0.25) -- (0.46875,0.3125) -- (0.48828125,0.3125) -- (0.48828125,0.375) -- (0.5078125,0.375) -- (0.5078125,0.4375) -- (0.52734375,0.4375) -- (0.52734375,0.5) -- (0.546875,0.5) -- (0.546875,0.4375) -- (0.56640625,0.4375) -- (0.56640625,0.5) -- (0.5859375,0.5) -- (0.5859375,-6.25e-07) -- cycle;\draw (0,0.5) -- (0,0) -- (0.5859375,0) -- (0.5859375,0.5); \end{tikzpicture}       &17            &180             &1800              \\
Gidney (2017) \cite{gidney2018halving}               &in    &Ripple Carry            &$n - 1$                         &$2n - 1$                       &$n$                        &\begin{tikzpicture}\fill[red] (0.0,-6.25e-07) -- (0.0,0.0) -- (0.0,0.0625) -- (2.48046875,0.0625) -- (2.48046875,0.0) -- (4.98046875,0.0) -- (4.98046875,-6.25e-07) -- cycle;\draw (0,0.5) -- (0,0) -- (4.98046875,0) -- (4.98046875,0.5); \end{tikzpicture}                                                                                                                                                                                                                                                                                                                                                                                                                                                                                                                                                                                                                                                                                                                                                                                                                                                                                                                                                                                                                                                                                             &1             &76              &6600              \\
Mogensen (2019) \cite{mogensen2019lookahead}         &in    &Carry Lookahead         &$6n - 4$                        &$6\lg n + 2$                   &$n - \lg n - 1$            &\begin{tikzpicture}\fill[red] (0.0,-6.25e-07) -- (0.0,0.0) -- (0.0,0.5) -- (0.0390625,0.5) -- (0.0390625,0.4375) -- (0.078125,0.4375) -- (0.078125,0.375) -- (0.1171875,0.375) -- (0.1171875,0.3125) -- (0.15625,0.3125) -- (0.15625,0.25) -- (0.1953125,0.25) -- (0.1953125,0.1875) -- (0.234375,0.1875) -- (0.234375,0.125) -- (0.3515625,0.125) -- (0.3515625,0.1875) -- (0.4296875,0.1875) -- (0.4296875,0.25) -- (0.5078125,0.25) -- (0.5078125,0.3125) -- (0.5859375,0.3125) -- (0.5859375,0.375) -- (0.6640625,0.375) -- (0.6640625,0.4375) -- (0.8203125,0.4375) -- (0.8203125,0.375) -- (0.8984375,0.375) -- (0.8984375,0.3125) -- (0.9765625,0.3125) -- (0.9765625,0.25) -- (1.0546875,0.25) -- (1.0546875,0.1875) -- (1.1328125,0.1875) -- (1.1328125,0.125) -- (1.25,0.125) -- (1.25,0.1875) -- (1.2890625,0.1875) -- (1.2890625,0.25) -- (1.328125,0.25) -- (1.328125,0.3125) -- (1.3671875,0.3125) -- (1.3671875,0.375) -- (1.40625,0.375) -- (1.40625,0.4375) -- (1.4453125,0.4375) -- (1.4453125,0.5) -- (1.484375,0.5) -- (1.484375,-6.25e-07) -- cycle;\draw (0,0.5) -- (0,0) -- (1.484375,0) -- (1.484375,0.5); \end{tikzpicture}                                                                                                                                                                                                     &14            &140             &1500              \\
Thapliyal et al (2020) \cite{thapliyal2020lookahead} &in    &Carry Lookahead         &$7n$*                           &$4\lg n + O(1)$*               &$2n$*                      &\begin{tikzpicture}\fill[red] (0.0,-6.25e-07) -- (0.0,0.0) -- (0.0,0.5) -- (0.01953125,0.5) -- (0.01953125,0.4375) -- (0.0390625,0.4375) -- (0.0390625,0.5) -- (0.05859375,0.5) -- (0.05859375,0.4375) -- (0.078125,0.4375) -- (0.078125,0.375) -- (0.09765625,0.375) -- (0.09765625,0.3125) -- (0.1171875,0.3125) -- (0.1171875,0.25) -- (0.13671875,0.25) -- (0.13671875,0.125) -- (0.17578125,0.125) -- (0.17578125,0.1875) -- (0.1953125,0.1875) -- (0.1953125,0.25) -- (0.21484375,0.25) -- (0.21484375,0.3125) -- (0.234375,0.3125) -- (0.234375,0.375) -- (0.25390625,0.375) -- (0.25390625,0.4375) -- (0.2734375,0.4375) -- (0.2734375,0.0) -- (0.3125,0.0) -- (0.3125,0.4375) -- (0.33203125,0.4375) -- (0.33203125,0.375) -- (0.3515625,0.375) -- (0.3515625,0.3125) -- (0.37109375,0.3125) -- (0.37109375,0.25) -- (0.390625,0.25) -- (0.390625,0.1875) -- (0.41015625,0.1875) -- (0.41015625,0.125) -- (0.44921875,0.125) -- (0.44921875,0.25) -- (0.46875,0.25) -- (0.46875,0.3125) -- (0.48828125,0.3125) -- (0.48828125,0.375) -- (0.5078125,0.375) -- (0.5078125,0.4375) -- (0.52734375,0.4375) -- (0.52734375,0.5) -- (0.546875,0.5) -- (0.546875,0.4375) -- (0.56640625,0.4375) -- (0.56640625,0.5) -- (0.5859375,0.5) -- (0.5859375,-6.25e-07) -- cycle;\draw (0,0.5) -- (0,0) -- (0.5859375,0) -- (0.5859375,0.5); \end{tikzpicture} &13            &130             &1300              \\
(this paper) (2020)                                  &in    &Blocksize=$\sqrt{n}$    &$5n + 4\sqrt{n} + O(1)$         &$6\sqrt{n} + 2\lg n + O(1)$    &$2n + 3\sqrt{n} + O(1)$    &\begin{tikzpicture}\fill[red] (0.0,-6.25e-07) -- (0.0,0.0) -- (0.0,0.375) -- (0.234375,0.375) -- (0.234375,0.3125) -- (0.25390625,0.3125) -- (0.25390625,0.25) -- (0.2734375,0.25) -- (0.2734375,0.1875) -- (0.29296875,0.1875) -- (0.29296875,0.25) -- (0.3125,0.25) -- (0.3125,0.0) -- (0.33203125,0.0) -- (0.33203125,0.3125) -- (0.56640625,0.3125) -- (0.56640625,0.0) -- (0.625,0.0) -- (0.625,0.25) -- (0.64453125,0.25) -- (0.64453125,0.3125) -- (0.6640625,0.3125) -- (0.6640625,0.0) -- (0.8984375,0.0) -- (0.8984375,0.375) -- (1.1328125,0.375) -- (1.1328125,0.3125) -- (1.15234375,0.3125) -- (1.15234375,0.1875) -- (1.171875,0.1875) -- (1.171875,0.125) -- (1.19140625,0.125) -- (1.19140625,0.25) -- (1.2109375,0.25) -- (1.2109375,0.0) -- (1.5234375,0.0) -- (1.5234375,0.1875) -- (1.54296875,0.1875) -- (1.54296875,0.3125) -- (1.5625,0.3125) -- (1.5625,0.0) -- (1.796875,0.0) -- (1.796875,-6.25e-07) -- cycle;\draw (0,0.5) -- (0,0) -- (1.796875,0) -- (1.796875,0.5); \end{tikzpicture}                                                                                                                                                                                                                                                                                                                                     &10            &96              &950               \\
(this paper) (2020)                                  &in    &Blocksize=b             &$5n - 4b + 8\frac{n}{b} + O(1)$ &$6b + 4\lg \frac{n}{b} + O(1)$ &$2n + 3\frac{n}{b} + O(1)$ &\begin{tikzpicture}\fill[red] (0.0,-6.25e-07) -- (0.0,0.0) -- (0.0,0.375) -- (0.1953125,0.375) -- (0.1953125,0.3125) -- (0.21484375,0.3125) -- (0.21484375,0.1875) -- (0.234375,0.1875) -- (0.234375,0.125) -- (0.25390625,0.125) -- (0.25390625,0.25) -- (0.2734375,0.25) -- (0.2734375,0.0) -- (0.29296875,0.0) -- (0.29296875,0.3125) -- (0.48828125,0.3125) -- (0.48828125,0.0) -- (0.546875,0.0) -- (0.546875,0.1875) -- (0.56640625,0.1875) -- (0.56640625,0.3125) -- (0.5859375,0.3125) -- (0.5859375,0.0) -- (0.78125,0.0) -- (0.78125,0.375) -- (0.9765625,0.375) -- (0.9765625,0.3125) -- (0.99609375,0.3125) -- (0.99609375,0.1875) -- (1.015625,0.1875) -- (1.015625,0.125) -- (1.03515625,0.125) -- (1.03515625,0.25) -- (1.0546875,0.25) -- (1.0546875,0.0) -- (1.328125,0.0) -- (1.328125,0.1875) -- (1.34765625,0.1875) -- (1.34765625,0.3125) -- (1.3671875,0.3125) -- (1.3671875,0.0) -- (1.5625,0.0) -- (1.5625,-6.25e-07) -- cycle;\draw (0,0.5) -- (0,0) -- (1.5625,0) -- (1.5625,0.5); \end{tikzpicture}                                                                                                                                                                                                                                                                                                                           &9             &95              &950               \\

\hline
Gossett (1998) \cite{gossett1998carrysave}           &out   &Carry Save (avg over n) &$4n$                            &$2$                            &$n^2 - 2n$                 &\begin{tikzpicture}\fill[red] (0.0,-6.25e-07) -- (0.0,0.0) -- (0.0,0.5625) -- (0.0390625,0.5625) -- (0.0390625,-6.25e-07) -- cycle;\draw (0,0.5) -- (0,0) -- (0.0390625,0) -- (0.0390625,0.5); \end{tikzpicture}                                                                                                                                                                                                                                                                                                                                                                                                                                                                                                                                                                                                                                                                                                                                                                                                                                                                                                                                                                                                                                                                                                                                         &71            &6600            &660000            \\
Draper et al. (2004) \cite{draper2004lookaheadadder} &out   &Carry Lookahead         &$5n - 3\lg n - 4$               &$2\lg n + 3$                   &$n - \lg n$                &\begin{tikzpicture}\fill[red] (0.0,-6.25e-07) -- (0.0,0.0) -- (0.0,0.5) -- (0.01953125,0.5) -- (0.01953125,0.4375) -- (0.0390625,0.4375) -- (0.0390625,0.5) -- (0.05859375,0.5) -- (0.05859375,0.4375) -- (0.078125,0.4375) -- (0.078125,0.375) -- (0.09765625,0.375) -- (0.09765625,0.3125) -- (0.1171875,0.3125) -- (0.1171875,0.25) -- (0.13671875,0.25) -- (0.13671875,0.125) -- (0.17578125,0.125) -- (0.17578125,0.25) -- (0.1953125,0.25) -- (0.1953125,0.3125) -- (0.21484375,0.3125) -- (0.21484375,0.375) -- (0.234375,0.375) -- (0.234375,0.4375) -- (0.25390625,0.4375) -- (0.25390625,0.5) -- (0.2734375,0.5) -- (0.2734375,0.4375) -- (0.29296875,0.4375) -- (0.29296875,-6.25e-07) -- cycle;\draw (0,0.5) -- (0,0) -- (0.29296875,0) -- (0.29296875,0.5); \end{tikzpicture}                                                                                                                                                                                                                                                                                                                                                                                                                                                                                                                                                               &8             &91              &920               \\
Gidney (2017) \cite{gidney2018halving}               &out   &Ripple Carry            &$n - 1$                         &$n - 1$                        &$1$                        &\begin{tikzpicture}\fill[red] (0.0,-6.25e-07) -- (0.0,0.0) -- (0.0,0.0625) -- (2.48046875,0.0625) -- (2.48046875,-6.25e-07) -- cycle;\draw (0,0.5) -- (0,0) -- (2.48046875,0) -- (2.48046875,0.5); \end{tikzpicture}                                                                                                                                                                                                                                                                                                                                                                                                                                                                                                                                                                                                                                                                                                                                                                                                                                                                                                                                                                                                                                                                                                                                     &1             &41              &3100              \\
Mogensen (2019) \cite{mogensen2019lookahead}         &out   &Carry Lookahead         &$12n - 8$                       &$12\lg n + 4$                  &$n - \lg n - 1$            &\begin{tikzpicture}\fill[red] (0.0,-6.25e-07) -- (0.0,0.0) -- (0.0,0.5) -- (0.0390625,0.5) -- (0.0390625,0.4375) -- (0.078125,0.4375) -- (0.078125,0.375) -- (0.1171875,0.375) -- (0.1171875,0.3125) -- (0.15625,0.3125) -- (0.15625,0.25) -- (0.1953125,0.25) -- (0.1953125,0.1875) -- (0.234375,0.1875) -- (0.234375,0.125) -- (0.3515625,0.125) -- (0.3515625,0.1875) -- (0.4296875,0.1875) -- (0.4296875,0.25) -- (0.5078125,0.25) -- (0.5078125,0.3125) -- (0.5859375,0.3125) -- (0.5859375,0.375) -- (0.6640625,0.375) -- (0.6640625,0.4375) -- (0.7421875,0.4375) -- (0.7421875,-6.25e-07) -- cycle;\draw (0,0.5) -- (0,0) -- (0.7421875,0) -- (0.7421875,0.5); \end{tikzpicture}                                                                                                                                                                                                                                                                                                                                                                                                                                                                                                                                                                                                                                                                 &19            &190             &1900              \\
Thapliyal et al (2020) \cite{thapliyal2020lookahead} &out   &Carry Lookahead         &$4n$                            &$2\lg n + O(1)$                &$n$*                       &\begin{tikzpicture}\fill[red] (0.0,-6.25e-07) -- (0.0,0.0) -- (0.0,0.5) -- (0.01953125,0.5) -- (0.01953125,0.4375) -- (0.0390625,0.4375) -- (0.0390625,0.5) -- (0.05859375,0.5) -- (0.05859375,0.4375) -- (0.078125,0.4375) -- (0.078125,0.375) -- (0.09765625,0.375) -- (0.09765625,0.3125) -- (0.1171875,0.3125) -- (0.1171875,0.25) -- (0.13671875,0.25) -- (0.13671875,0.125) -- (0.17578125,0.125) -- (0.17578125,0.1875) -- (0.1953125,0.1875) -- (0.1953125,0.25) -- (0.21484375,0.25) -- (0.21484375,0.3125) -- (0.234375,0.3125) -- (0.234375,0.375) -- (0.25390625,0.375) -- (0.25390625,0.4375) -- (0.2734375,0.4375) -- (0.2734375,0.0) -- (0.29296875,0.0) -- (0.29296875,-6.25e-07) -- cycle;\draw (0,0.5) -- (0,0) -- (0.29296875,0) -- (0.29296875,0.5); \end{tikzpicture}                                                                                                                                                                                                                                                                                                                                                                                                                                                                                                                                                               &7             &73              &730               \\
(this paper) (2020)                                  &out   &Blocksize=$\sqrt{n}$    &$3n + 3\sqrt{n} + O(1)$         &$3\sqrt{n} + \lg n + O(1)$     &$2n + 3\sqrt{n} + O(1)$    &\begin{tikzpicture}\fill[red] (0.0,-6.25e-07) -- (0.0,0.0) -- (0.0,0.375) -- (0.234375,0.375) -- (0.234375,0.3125) -- (0.25390625,0.3125) -- (0.25390625,0.25) -- (0.2734375,0.25) -- (0.2734375,0.1875) -- (0.29296875,0.1875) -- (0.29296875,0.25) -- (0.3125,0.25) -- (0.3125,0.0) -- (0.33203125,0.0) -- (0.33203125,0.3125) -- (0.56640625,0.3125) -- (0.56640625,0.0) -- (0.625,0.0) -- (0.625,0.25) -- (0.64453125,0.25) -- (0.64453125,0.3125) -- (0.6640625,0.3125) -- (0.6640625,0.0) -- (0.8984375,0.0) -- (0.8984375,-6.25e-07) -- cycle;\draw (0,0.5) -- (0,0) -- (0.8984375,0) -- (0.8984375,0.5); \end{tikzpicture}                                                                                                                                                                                                                                                                                                                                                                                                                                                                                                                                                                                                                                                                                                                       &6             &63              &610               \\
(this paper) (2020)                                  &out   &Blocksize=b             &$3n - 2b + 5\frac{n}{b} + O(1)$ &$3b + 2\lg \frac{n}{b} + O(1)$ &$2n + 3\frac{n}{b} + O(1)$ &\begin{tikzpicture}\fill[red] (0.0,-6.25e-07) -- (0.0,0.0) -- (0.0,0.375) -- (0.1953125,0.375) -- (0.1953125,0.3125) -- (0.21484375,0.3125) -- (0.21484375,0.1875) -- (0.234375,0.1875) -- (0.234375,0.125) -- (0.25390625,0.125) -- (0.25390625,0.25) -- (0.2734375,0.25) -- (0.2734375,0.0) -- (0.29296875,0.0) -- (0.29296875,0.3125) -- (0.48828125,0.3125) -- (0.48828125,0.0) -- (0.546875,0.0) -- (0.546875,0.1875) -- (0.56640625,0.1875) -- (0.56640625,0.3125) -- (0.5859375,0.3125) -- (0.5859375,0.0) -- (0.78125,0.0) -- (0.78125,-6.25e-07) -- cycle;\draw (0,0.5) -- (0,0) -- (0.78125,0) -- (0.78125,0.5); \end{tikzpicture}                                                                                                                                                                                                                                                                                                                                                                                                                                                                                                                                                                                                                                                                                                             &6             &62              &610               \\
\end{tabular}

}
    \caption{Comparison of various adders.
    $V(n,f)$ is the estimated logical qubit seconds needed to execute an $n$-bit adder using at most $f$ magic state factories (see \sec{estimate}).
    An asterisk (*) means value differs from original paper (see \app{correction}).
    Generated by \texttt{generate\_figures.py}.
    }
    \label{tbl:comparison}
\end{table}

\begin{figure}[H]
    \centering
    \minipage{0.5\textwidth}
    \resizebox{\linewidth}{!}{
    \includegraphics{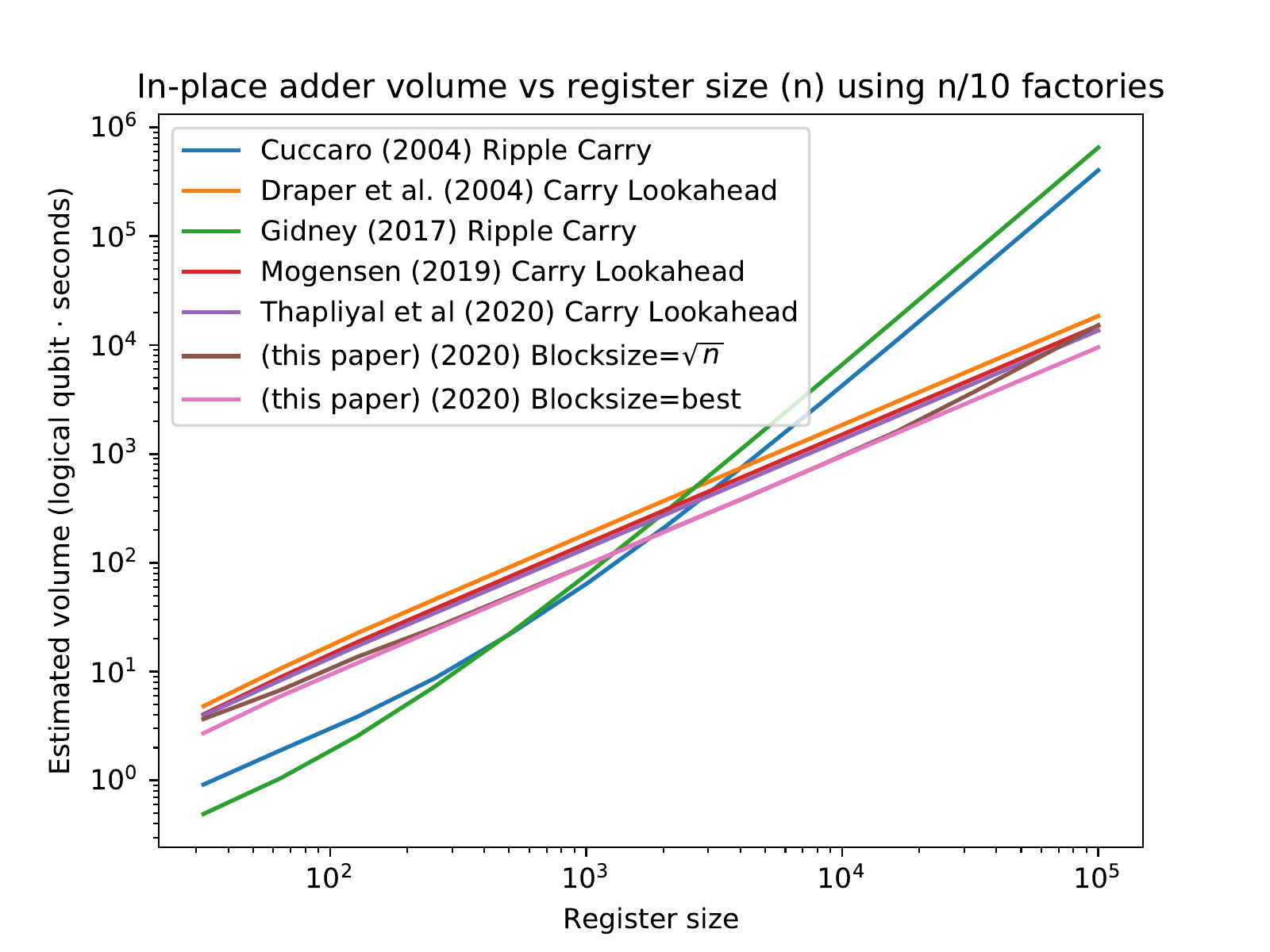}
    }
    \endminipage
    \minipage{0.5\textwidth}
    \resizebox{\linewidth}{!}{
    \includegraphics{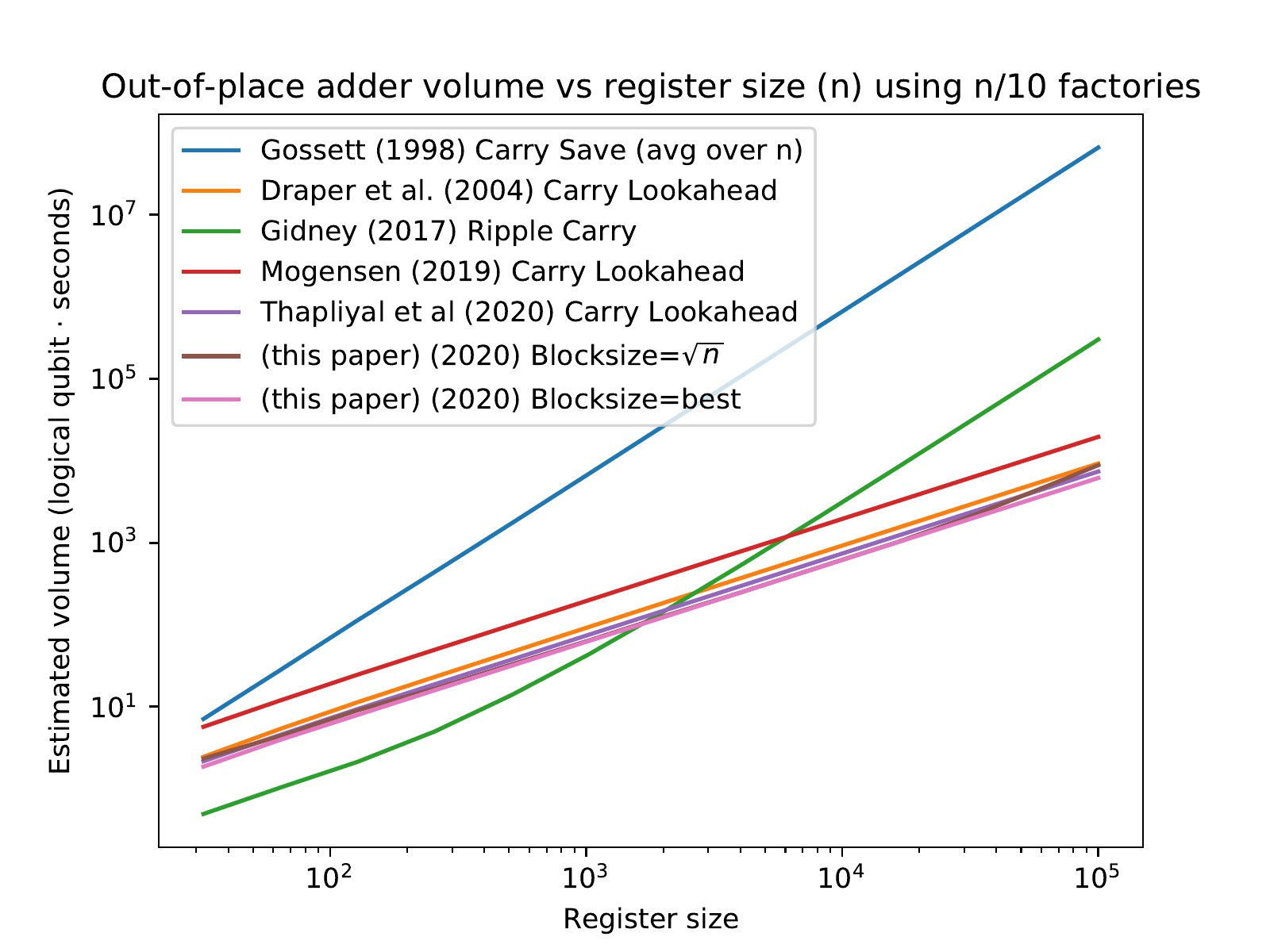}
    }
    \endminipage
    \caption{
        Log-log plot of adder spacetime volume versus register size. Assumes the maximum factory count is one tenth the register size, that factories have a footprint of 72 logical qubits, that factories produce a magic state every 165 microseconds, and that the control system reaction time is 10 microseconds.
        Generated by \texttt{generate\_figures.py}.
    }
    \label{fig:size_versus_volume}
\end{figure}

\begin{figure}[H]
    \centering
    \resizebox{\linewidth}{!}{
    \includegraphics{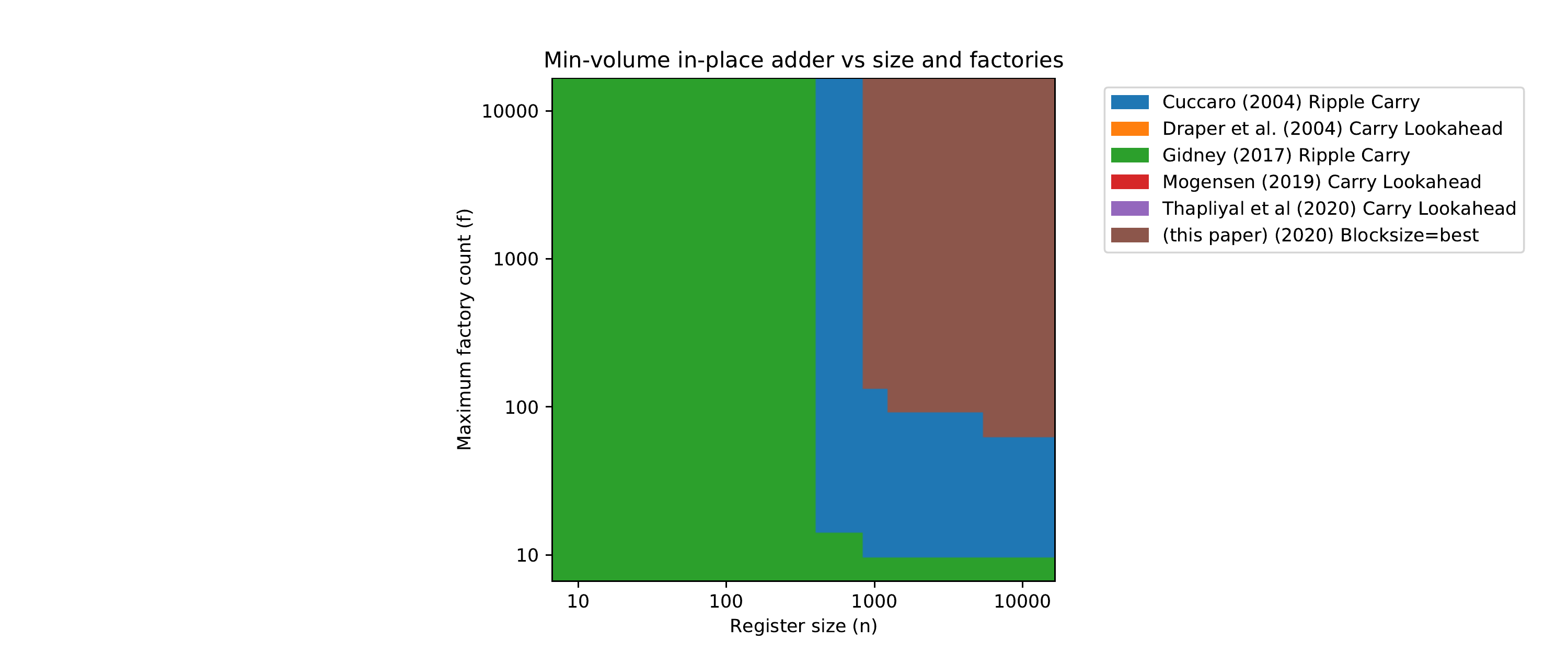}
    }
    \caption{
        Lowest volume in place adder at various register sizes and maximum factory counts.
        Assumes that factories have a footprint of 72 logical qubits, that factories produce a magic state every 165 microseconds, and that the control system reaction time is 10 microseconds.
        Generated by \texttt{generate\_figures.py}.
    }
    \label{fig:minif}
\end{figure}

\begin{figure}[H]
    \centering
    \resizebox{\linewidth}{!}{
    \includegraphics{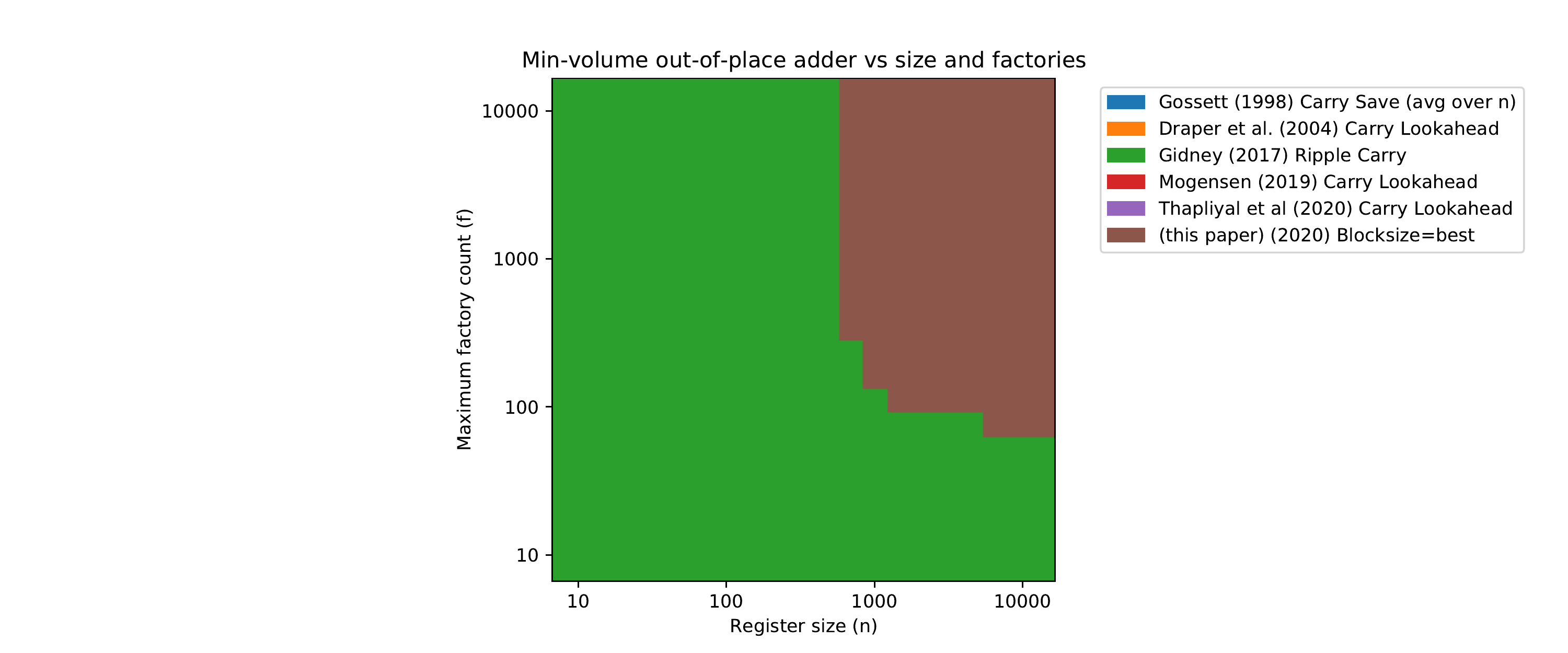}
    }
    \caption{
        Lowest volume out of place adder at various register sizes and maximum factory counts.
        Assumes that factories have a footprint of 72 logical qubits, that factories produce a magic state every 165 microseconds, and that the control system reaction time is 10 microseconds.
        Generated by \texttt{generate\_figures.py}.
    }
    \label{fig:minof}
\end{figure}

\begin{figure}[H]
    \centering
    \resizebox{\linewidth}{!}{
    \includegraphics{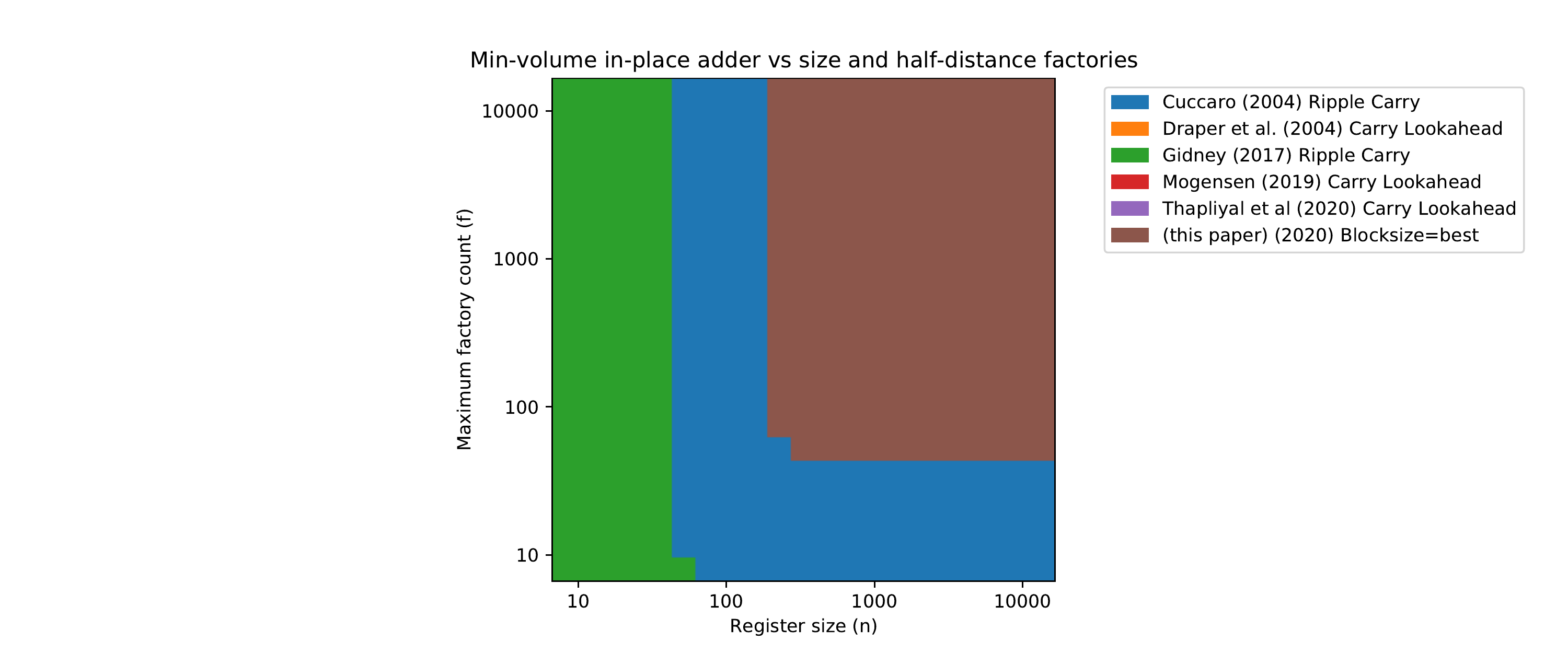}
    }
    \resizebox{\linewidth}{!}{
    \includegraphics{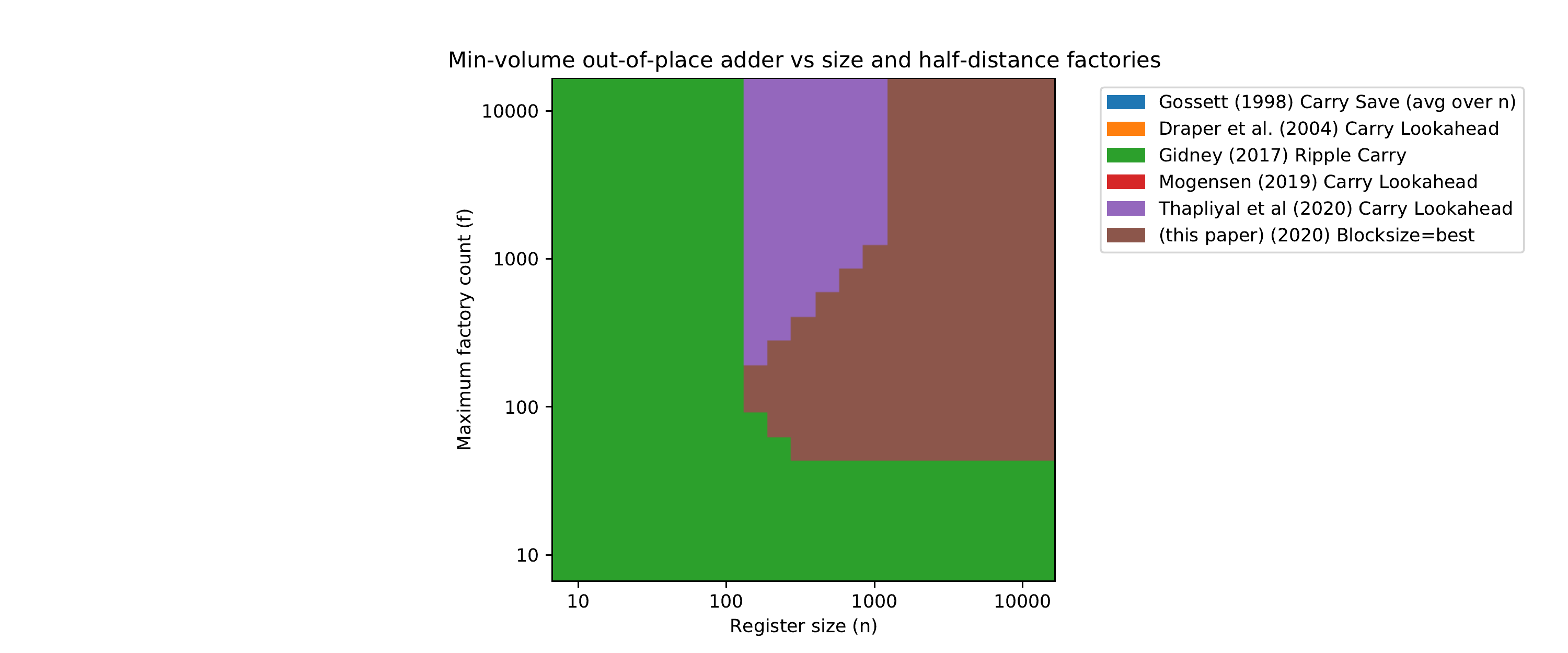}
    }
    \caption{
        Lowest volume in place and out of place adders at various register sizes and maximum factory counts, assuming improved magic state factories.
        Assumes that factories have a footprint of 18 logical qubits, that factories produce a magic state every 82.5 microseconds, and that the control system reaction time is 10 microseconds.
        Generated by \texttt{generate\_figures.py}.
    }
    \label{fig:minioh}
\end{figure}

Although the example we just gave is in terms of the execution time of a circuit, focusing on just time can be misleading.
In this paper, we're going to focus on the spacetime volume consumed by a circuit instead.
This choice is based on the idea that any volume not used for one thing can be repurposed for some other useful task (like magic state distillation or temporary storage).
The topological nature of surface code computations, which allows for deformations and rotations through spacetime, makes this assumption surprisingly robust.
That being said, beware that there are regimes where focusing on spacetime volume is the wrong thing to do.
For example, in a world where space is limitless, making magic state factories effectively free, time is king and low depth adders are almost trivially better than ripple carry adders.
But that is not the parameter regime we are interested in, so we will focus on volume.

Our goal in this paper is to reduce and quantify the spacetime overhead implicit in using a low depth adder.
In \sec{block}, we present a lookahead adder that parallelizes over blocks of size $b$ instead of over bits.
In \sec{estimate}, we estimate the spacetime volume of this adder and previous adders in order to understand the parameter regimes where various adders dominate.
Finally, in \sec{conclusion}, we summarize our results and give some additional caveats.

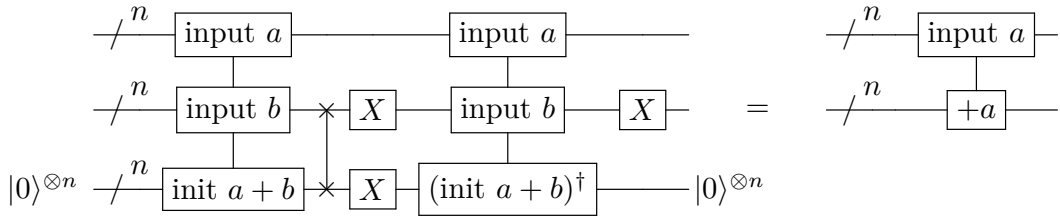
\begin{figure}
\centering
\resizebox{0.85\linewidth}{!}{
\Qcircuit @R=1em @C=0.75em {
\\
&{/} \qw& \ustick{n}\qw&\gate{\text{input }a}     &\qw       &\qw       &\gate{\text{input }a}              &\qw       &\qw& & & & & & &{/} \qw& \ustick{n}\qw&\qw&\gate{\text{input }a}&\qw&\\
&{/} \qw& \ustick{n}\qw&\gate{\text{input }b} \qwx&\qswap    &\gate{X}  &\gate{\text{input }b}          \qwx&\gate{X}  &\qw& & &=& & & &{/} \qw& \ustick{n}\qw&\qw&\gate{\text{+}a}\qwx&\qw&\\
\lstick{|0\rangle^{\otimes n}}&{/} \qw& \ustick{n}\qw&\gate{\text{init }a+b}\qwx&\qswap\qwx&\gate{X}  &\gate{(\text{init }a+b)^\dagger}\qwx&\qw       &\qw&&&&\lstick{|0\rangle^{\otimes n}}& & & & & & & & \\
\\
}
}
    \caption{
        Converting an out of place adder into an in place adder by running the out of place adder forwards and then backwards, with a few additional swap and Pauli operations.
        Swap and Pauli operations can be tracked within the classical control system instead of actually being applied to the qubits.
    }
    \label{fig:oop2ip}
\end{figure}

\section{Block Lookahead Adder}
\label{sec:block}

To give a sense of how a block lookahead adder works, we will first explain a simple case: the two block adder.
The two block adder split an $n$-bit addition problem into two $n/2$-bit blocks.
It starts by performing three ripple carry additions in parallel: adding the low blocks of each input using no carry input, adding the high blocks of each input using no carry input, and adding the high blocks of each input using a set carry input.
As soon as the low block addition produces a carry output, this carry output is used to decide which of the two high block addition results to keep.
With the correct high block addition written into the output register, the high block additions become garbage and are uncomputed.

Here is pseudo-code describing the two block adder:

\begin{python}
    # Parallel ripple-carry adders.
    let out_low = a_low + b_low carryinto carry_out
    let case0 = a_high + b_high
    let case1 = a_high + b_high + 1
    # Choose high half using carry_out from low half.
    let out_high = case0 if carry_out else case1
    # Uncompute intermediate values in parallel.
    del carry_out
    del case1
    del case0
\end{python}

The ancillary file \texttt{src/adder\_two\_block.qs} has Q\# code implementing this adder.

We can generalize the two block adder into an adder with $n/b$ blocks of size $b$.
For each block, the block lookahead adder computes both the carry-in-cleared and carry-in-set cases.
It then converts the carry-out values from these local additions into local ``propagate" and ``generate" signals.
These can be fed into the same subroutine used by a carry lookahead adder to mutate the local ``generate" signals into propagated global ``generate" signals.
Once the global signals are known, they are used to choose for each block whether to keep the output from the carry-in-cleared or carry-in-set case.
With that done, the final sum is ready and all garbage is uncomputed.

Parallelizing over blocks in this way, instead of over bits, changes the Toffoli count of out of place addition from $4n$ to $3n + 5n/b$.
This is beneficial as long as $b > 5$ and the number of magic state factories is a bottleneck.
The tradeoff for the improved Toffoli count is that the reaction depth increases from $2 \lg n + O(1)$ to $3 b + 2 \lg(n/b)$, and the workspace increases from $n$ to $2n + 3n/b$.
The workspace increase looks very large, but it's important to remember that despite using more workspace we're requiring fewer magic state factories (to run at a given speed).
Whether or not these costs balance out is the subject of the next section.

One complication in the construction of the block lookahead adder is that the carry lookahead subroutine overwrites local carry values with propagated carry values (instead of storing the propagated carry values in separate qubits), but the local carry values are needed when uncomputing other garbage.
We could fix this by uncomputing the carry propagation subroutine after copying its output to another register, but we don't want to do that because it increases the Toffoli count and also at small block sizes that subroutine is the dominant contributor to the reaction depth.
Instead, each updated carry value is cleared using the fact that they are equal to the xor of the output sum and the two input summands at some bit position.
The original carry values can then be restored by repeating the last step of the initial ripple carry sums.

So far we have only described an out of place block lookahead adder.
To convert it into an in place adder, we use the adder-independent construction shown in \fig{oop2ip}.

We implemented the block lookahead adder in Q\#.
The source code can be found in \app{blockadder}, or in the paper's ancillary files.
The top level method is \texttt{init\_sum\_using\_blocks} in the ancillary file \\\texttt{src/adder\_blocks.qs}.
For the carry propagation subroutine, we implemented the T-optimized out of place adder described by Thapliyal et al in \cite{thapliyal2020lookahead}.
The ripple carry adder we used is the out of place adder from \cite{gidney2018halving}.

\section{Spacetime Volume Comparison}
\label{sec:estimate}

In a tiny quantum computer that can barely support a single magic state factory, ripple carry adders will always be best.
Conversely, in a huge quantum computer where space is too cheap to meter, low depth adders will dominate.
In this section our goal is to understand the intermediate regime between these two extremes.

Before we discuss our methods, we caution the reader to think carefully when extrapolating our results to other parameter regimes.
For example, you might expect that using slower physical qubits (e.g. switching from superconducting qubits to ion trap qubits) would increase the need for low depth adders to compensate for the slowdown.
Actually, the reverse occurs.
Architectures with slower physical qubits have a larger ratio $r$ between the reaction time of the control system and the period of magic state factories.
They can do more control system round trips per magic state, meaning reaction-limited computation can feed the outputs of more factories into a single serial circuit.
Basically, slowing down the qubits doesn't slow down the ripple carry adders, because the ripply carry process is mainly limited by the control system's reaction time (until the physical qubits get so slow that even physical measurement is slower than the control system).
Instead, using slower qubits increases the relative cost of magic state factories.
So, at least in terms of spacetime volume, slowing down physical qubits actually favors serial constructions that use fewer Toffoli gates (such as ripple carry adders).

To compare adders, we started by looking up or computing their Toffoli count, reaction depth, workspace, and Toffoli consumption rates over time.
These values are listed in \tbl{comparison}.
The Toffoli count is the number of Toffoli magic states that the circuit consumes when executed.
The reaction depth is the deepest chain of sequential adaptive measurements that occur in the circuit (keeping in mind that in the surface code every Toffoli gate, AND computation, and AND uncomputation involves an adaptive measurement).
The workspace is the number of additional logical qubits needed in the abstract circuit model to run the circuit (i.e. not counting logical qubits used for magic state factories or for storing the input/output).

With the adder data collected, we wrote code to estimate the spacetime volume of an adder given a specified reaction time, factory footprint, and factory period.
The code can be found in the \texttt{vol} method of the \texttt{Adder} class in the ancillary file \texttt{generate\_figures.py}.
The code adds together spacetime estimates for the distillation of Toffoli states needed by the circuit, for storage of the circuit's input and output data, for storage of temporary qubits allocated during the execution of the circuit, and for buffering magic states during low-Toffoli-consumption periods in the circuit.
If an adder needed fewer factories than the maximum, we did not count the unused factories against its spacetime volume (i.e. we assumed they would be used for something else instead).

We used the volume estimation code to generate several figures.
\fig{size_versus_volume} plots the estimated spacetime volume of an adder versus the register size of the addition being performed, assuming the maximum factory count is 10\% of the register size.
Note how, in this figure, the best adders are ripple-carry adders until register sizes are in the thousands.
In \fig{minif} and \fig{minof} we plotted 2d heatmaps showing the in place and out of place adders with the best estimated spacetime volume, for various register sizes and maximum factory counts.
Once again the best adders are ripple-carry adders until register sizes are in the thousands, but this plot also shows that ripple carry adders dominate until factory counts are nearly 100.
\fig{minioh} also has 2d heatmaps showing the adder with the best estimated spacetime volume, but assumes improved factories using half the code distance (i.e. a quarter the footprint area and half the duration).
This moves the transition from ripple-carry adders to lookahead adders downwards, to register sizes in the hundreds of qubits and factory counts around 50.

\section{Conclusion}
\label{sec:conclusion}

In this paper we presented a block lookahead adder with a lower Toffoli count than in previous work, at the cost of a higher depth and more workspace.
We implemented these adders in Q\#, which allowed us to verify that they computed the correct result (see ancillary file \texttt{test/adder\_tests.qs}) while using at most the stated number of Toffolis and workspace qubits (see ancillary file \\\texttt{test/budget\_tests.qs}).
Unfortunately, Q\# is not currently able to verify the reaction depth.
We wrote Python code to estimate the spacetime volume of these adders, and previous adders, and compared them over a range of register sizes and maximum factory counts.
We found that our block lookahead adder has lower spacetime volume than previous low depth adders.
However, we also demonstrated that ripple carry adders have the best estimated spacetime volumes for register sizes up to 200 and for maximum factory counts up to 50.
This is a surprising contrast to the classical world, where carry lookahead adders are used even for 8 bit additions \cite{shirriff2020reverseengineer8008}.

An important type of adder that we didn't consider in this paper is adders that operate on representations of an integer other than 2s complement, like Draper's frequency-space adder \cite{draper2000qftaddition}.
These adders can't be used in all situations, due to format limitations and conversion costs, but they're sometimes the best choice.
For example, Shor's factoring algorithm \cite{shor1994algorithms} does arithmetic on registers whose sizes are in the thousands, making it a candidate for lookahead adders.
But, Shor's algorithm is compatible with an integer format that splits the addition problem into (an exponentially good approximation of) independent pieces \cite{gidney2019approximate,gidney2019factor}.
It's more efficient to apply ripple carry adders to the pieces than to apply lookahead adders to the whole register.

Overall, we believe that a clear conclusion of our numerics is that lookahead adders are rarely the best choice of quantum adder.
At least, not in the cost model we have chosen.
There is too much competition from below in the form of ripple carry adders, and from above in the form of other alternatives for parallelization (such as reaction limited computation and carry runways).
We can only hope that, as theoretical and engineering advances are made, and the projected cost of quantum fault tolerance steadily decreases, hindsight will show that this conclusion was quaintly naive.

\bibliographystyle{plain}
\bibliography{references}

\appendix

\section{Corrections in Table 1}
\label{app:correction}

Forewarned by \cite{oonishi2020efficient} that the resource counts of the T-optimized in place adder in \cite{thapliyal2020lookahead} were reported incorrectly, we double checked the values.
We counted the gates used during each step and also inspected the behavior of our Q\# implementation of their adder (which we use as a subroutine).
Our results agree with \cite{oonishi2020efficient}, and we report these corrected values in \tbl{comparison}.

\section{Q\# Source Code for Block Lookahead Adder}
\label{app:blockadder}

The following source code is also available as a Visual Studio Code project in the ancillary files of this paper.
Assuming Q\# is installed and configured correctly, an example addition can be run via ``\texttt{dotnet run --project src/adder.csproj}" or the unit tests can be run by invoking ``\texttt{dotnet test test/tests.csproj}".

This code has been slightly modified compared to the code in the ancillary files, so that it is self-contained and does not run off the page.

\begin{qsharp}
namespace BlockAdder {
    open Microsoft.Quantum.Arrays;
    open Microsoft.Quantum.Arithmetic;
    open Microsoft.Quantum.Bitwise;
    open Microsoft.Quantum.Canon;
    open Microsoft.Quantum.Convert;
    open Microsoft.Quantum.Intrinsic;
    open Microsoft.Quantum.Math;
    open Microsoft.Quantum.Random;

    /// Perform `out_c := a+b` in O(block_size + lg(n)) depth.
    /// Assumes Length(a) = Length(b) = Length(out_c)
    /// Assumes out_c is zero'd.
    operation init_sum_using_blocks(
            block_size: Int,
            a: LittleEndian,
            b: LittleEndian,
            out_c: LittleEndian) : Unit is Adj {
        if (Length(a!) <= block_size) {
            init_sum_using_ripple_carry(a, b, out_c);
        } else {
            _init_sum_using_blocks_helper(block_size, a, b, out_c);
        }
    }

    operation _init_sum_using_blocks_helper(
            block_size: Int,
            a: LittleEndian,
            b: LittleEndian,
            out_c: LittleEndian) : Unit is Adj {
        let a_blocks = Chunks(block_size, a!);
        let b_blocks = Chunks(block_size, b!);
        let c_blocks = Chunks(block_size, out_c!);
        let n = Length(a!);
        let m = Length(a_blocks);

        using ((carries_0, carries_1, mux_0, mux_1) = (
                Qubit[m], 
                Qubit[m], 
                Qubit[n - block_size], 
                Qubit[n - block_size])) {
            let case_blocks_0 = [new Qubit[0]] + Chunks(block_size, mux_0);
            let case_blocks_1 = [new Qubit[0]] + Chunks(block_size, mux_1);

            // Only one low block case. Compute in parallel with the high cases.
            init_sum_using_ripple_carry(
                LittleEndian(a_blocks[0]),
                LittleEndian(b_blocks[0]),
                LittleEndian(c_blocks[0] + [carries_0[0]]));
            within {
                // Set the carry-in bits of case_blocks_1.
                for (k in 1..Length(case_blocks_1)-1) {
                    X(case_blocks_1[k][0]);
                }

                // Compute carry-in-cleared and carry-in-set cases in parallel.
                for (k in 1..m-1) {
                    let stop = k == m - 1 ? -1 | 0;
                    mutable m0 = case_blocks_0[k] + [carries_0[k]][...stop];
                    mutable m1 = case_blocks_1[k] + [carries_1[k]][...stop];
                    for (t in [m0, m1]) {
                        init_sum_using_ripple_carry(
                            LittleEndian(a_blocks[k]),
                            LittleEndian(b_blocks[k]),
                            LittleEndian(t));
                    }
                }
                
                // Currently carries_0 is local `generate` signals.
                // Convert carries_1 into local `propagate` signals.
                for (k in 1..m-1) {
                    CNOT(carries_0[k], carries_1[k]);
                }
            } apply {
                // Determine propagated carries using carry-lookahead strategy.
                _prop_gen(carries_1[1...] + carries_1[...0], carries_0);

                // Use propagated carries to pick which blocks to keep.
                for (k in 1..m-1) {
                    init_choose(
                        carries_0[k - 1],
                        case_blocks_0[k],
                        case_blocks_1[k],
                        c_blocks[k]);
                }

                // Clear propagated carries.
                for (k in 1..m-1) {
                    CNOT(a_blocks[k][0], carries_0[k-1]);
                    CNOT(b_blocks[k][0], carries_0[k-1]);
                    CNOT(c_blocks[k][0], carries_0[k-1]);
                }

                // Restore carries_0 (except for carries_0[0] left zero).
                for (k in 1..m-2) {
                    let af = Last(a_blocks[k]);
                    let bf = Last(b_blocks[k]);
                    let cf = Last(case_blocks_0[k]);
                    let tf = carries_0[k];
                    within {
                        X(cf);
                        CNOT(af, bf);
                        CNOT(af, cf);
                    } apply {
                        init_and(bf, cf, tf);
                        CNOT(af, tf);
                    }
                }
            }
        }
    }

    /// Performs `out_sum := a+b` in linear depth.
    /// Supports carry-in via `out_c[0]` and carry-out via longer `out_c`.
    /// Reference: https://arxiv.org/abs/1709.06648
    operation init_sum_using_ripple_carry(
            a: LittleEndian,
            b: LittleEndian,
            out_sum: LittleEndian) : Unit is Adj {
        let n = Length(a!);
        for (k in 0..Length(out_sum!)-2) {
            init_full_adder_step(a![k], b![k], out_sum![k], out_sum![k+1]);
        }
        if (n > 0 and n == Length(out_sum!)) {
            CNOT(a![n - 1], out_sum![n - 1]);
            CNOT(b![n - 1], out_sum![n - 1]);
        }
    }

    // Performs `2*out_2 + out_1 := a + b + c`.
    operation init_full_adder_step(
            a: Qubit,
            b: Qubit,
            mut_c_to_out_1: Qubit,
            out_2: Qubit) : Unit is Adj {
        CNOT(a, b);
        CNOT(a, mut_c_to_out_1);
        init_and(b, mut_c_to_out_1, out_2);
        CNOT(a, b);
        CNOT(a, out_2);
        CNOT(b, mut_c_to_out_1);
    }

    /// Performs `target := a &&& b`.
    operation init_and(a: Qubit, b: Qubit, target: Qubit) : Unit is Adj {
        body(...) {
            CCNOT(a, b, target);
        }
        adjoint(...) {
            // Uncomment this when using Toffoli simulator.
            CCNOT(a, b, target);
            // Uncomment this when using resource estimator.
            // H(target);
            // if (M(target) == One) {
            //     CZ(a, b);
            // }
        }
    }

    /// Finds the `propagate` qubit for the given range.
    function _range_p_storage(ps: Qubit[], range: Range) : Qubit {
        let start = RangeStart(range);
        let end = RangeEnd(range);
        if (end == start + 1) {
            return ps[start];
        }
        return ps[(Length(ps) + start + end) / 2];
    }

    /// Finds the `generate` qubit for the given range.
    /// Note qubits are re-used for multiple ranges.
    function _range_g_storage(out_c: Qubit[], range: Range) : Qubit {
        let start = RangeStart(range);
        let end = RangeEnd(range);
        if (end == start + 1) {
            return out_c[end];
        }
        mutable i = (start + end) / 2;
        for (v in FactorsOf2(i)-1..-1..0) {
            let m = 1 <<< v;
            if (i + m < Length(out_c)) {
                set i += m;
            }
        }
        return out_c[i + 1];
    }

    /// Propagates local `generate` values into global `generate` values.
    /// Reference: https://arxiv.org/abs/2004.01826
    /// Example:
    ///       propagates = ...111...1...1.....
    ///      mut_gs (in) = .....1...1......1..
    ///     mut_gs (out) = ..1111..11......1..
    operation _prop_gen(propagates: Qubit[], mut_gs: Qubit[]) : Unit is Adj {
        let n = Length(propagates);
        using (workspace = Qubit[n]) {
            let p = _range_p_storage(propagates + workspace, _);
            let g = _range_g_storage(mut_gs, _);
            for (step in PowersOfTwoBelow(n)) {
                for (i in 0..2*step..n) {
                    let j = i + step;
                    let k = j + step;
                    if (k < n) {
                        init_and(p(i..j), p(j..k), p(i..k));
                        CCNOT(g(i..j), p(j..k), g(i..k));
                    }
                }
            }
            for (step in (PowersOfTwoBelow(n))[...-1...]) {
                for (i in 0..2*step..n) {
                    let j = i + step;
                    let k = j + step;
                    if (k < n) {
                        Adjoint init_and(p(i..j), p(j..k), p(i..k));
                    }
                    if (j < n) {
                        CCNOT(g(i-1..i), p(i..j), g(i..j));
                    }
                }
            }
        }
    }

    // Returns the last item in an array.
    function Last<'T>(items: 'T[]) : 'T {
        return items[Length(items) - 1];
    }

    function PowersOfTwoBelow(n: Int) : Int[] {
        mutable result = new Int[0];
        mutable k = 1;
        while (k < n) {
            set result += [k];
            set k <<<= 1;
        }
        return result;
    }

    // Determines how many times `n` is divisible by 2.
    function FactorsOf2(n: Int) : Int {
        mutable k = n;
        mutable r = 0;
        while (k != 0 and k % 2 == 0) {
            set k >>>= 1;
            set r += 1;
        }
        return r;
    }

    /// Performs `out_target := control ? option1 | option0`.
    operation init_choose(
            control: Qubit,
            option0: Qubit[],
            option1: Qubit[],
            out_target: Qubit[]) : Unit is Adj {
        let n = Length(option0);
        for (k in 0..n-1) {
            CNOT(option1[k], option0[k]);
            init_and(control, option0[k], out_target[k]);
            CNOT(option1[k], option0[k]);
            CNOT(option0[k], out_target[k]);
        }
    }
}
\end{qsharp}

\end{document}